\documentclass[aps,prl,notitlepage,floatfix,amsmath,amssymb]{revtex4-1}%

\usepackage{graphicx}
\usepackage{dcolumn}
\usepackage{bm}

\begin{document}

\title{Thermally Driven Imbibition and Drainage Induced by Terraced Nanostructures}

\author{Carlos E. Colosqui}
\email[]{carlos.colosqui@stonybrook.edu}

\author{Teng Teng}
\author{Amir M. Rahmani}

\affiliation{Department of Mechanical Engineering, Stony Brook University, Stony Brook, NY 11794, USA.}
%
%\date{\today}
%
%%%%%%%%%%%%%%%%%%%%%%%%%%%%%%%%%%%%%%%%%%%%%%%%%%%%%%%%%%%%%%%%%%%%%%%%%%%%%%%%%%%%%%%%%%%%%%%%%%%
%
% ABSTRACT
%
%%%%%%%%%%%%%%%%%%%%%%%%%%%%%%%%%%%%%%%%%%%%%%%%%%%%%%%%%%%%%%%%%%%%%%%%%%
%
\begin{abstract}
Theoretical analysis and fully atomistic molecular dynamics simulations reveal a Brownian ratchet mechanism by which thermal fluctuations drive the net displacement of immiscible liquids confined in channels or pores with micro- or nanoscale dimensions.
The thermally-driven displacement is induced by surface nanostructures with directional asymmetry and can occur against the direction of action of wetting or capillary forces.
Mean displacement rates in molecular dynamics simulations are predicted via analytical solution of a Smoluchowski diffusion equation for the position probability density.
The proposed physical mechanisms and derived analytical expressions can be applied to engineer surface nanostructures for controlling the dynamics of diverse wetting processes such as capillary filling, wicking, and imbibition in micro- or nanoscale systems. 
\end{abstract}
%
%
%\pacs{47.85.-g; 82.70.Dd; 47.61.Jd}
%
%\keywords{}
\maketitle
%
%
%%%%%%%%%%%%%%%%%%%%%%%%%%%%%%%%%%%%%%%%%%%%%%%%%%%%%%%%%%%%%%%%%%%%%%%%%%%%%%%%%%%%%%%%%%%%%%%%%%
%   INTRODUCTION: NANOFLUIDICS - WETTING DYNAMICS -THERMALLY-ACTIVATED PROCESSES
%%%%%%%%%%%%%%%%%%%%%%%%%%%%%%%%%%%%%%%%%%%%%%%%%%%%%%%%%%%%%%%%%%%%%%%%%%%%%%%%%%%%%%%%%%%%%%%%%%
%
\section{Introduction: Nanoscale Wetting and Brownian Ratchets}
Advances in nanofabrication and characterization techniques have enabled the engineering of nanostructured surfaces with geometric features as small as a few nanometers \cite{rosei2004,tawfick2012,checco2014robust}.
At nanoscales, the interplay between intermolecular forces, Brownian motion, and surface structure can give rise to complex interfacial phenomena that are challenging for the application of conventional, continuum-based and deterministic, models \cite{schoch2008,rauscher2008,bocquet2010,snoeijer2013,colosqui2013meso}.
For example, nanoscale surface structures can induce energy barriers that lead to wetting processes governed by thermally-activated transitions between metastable states \cite{davidovitch2005,kaz2012,kavousanakis2013,colosqui2013,razavi2014,duvivier2013toward}.
These thermally-activated transitions can result in directed transport of fluids and solutes when there is directional asymmetry of the energy barriers induced by the physicochemical structure of the confining surfaces \cite{chinappi2006,fu2007,zuo2009transport,sparreboom2010transport}.   
Analogous mechanisms for rectification of thermal motion into directed transport underlie fundamental biological processes such as selective charge transport in ion channels or translocation of proteins across cellular membranes.
Physical systems where thermal fluctuations are able to drive net directional motion, while performing work against ``load'' or resistance forces, are known as thermal ratchets or Brownian motors and have been extensively studied in the framework of statistical physics \cite{astumian1994fluctuation,reimann2002brownian,hanggi2009artificial}.
%
%

%%%%%%%%%%%%%%%%%%%%%%%%%%%%%%%%%%%%%%%%%%%%%%%%%%%%%%%%%%%%%%%%%%%%%%%%%%%%%%%%%%%%%%%%%%%%%%%%%%   
% RATCHETS
%%%%%%%%%%%%%%%%%%%%%%%%%%%%%%%%%%%%%%%%%%%%%%%%%%%%%%%%%%%%%%%%%%%%%%%%%%%%%%%%%%%%%%%%%%%%%%%%%%
Thermal ratchets can operate without thermal or chemical gradients provided that the system has not reached all necessary conditions for thermodynamic equilibrium \cite{reimann2002brownian,hanggi2009artificial}. 
A variety of novel nano/microfluidic devices perform as thermal ratchets to accomplish the handling, separation, and detection of diverse solutes (e.g., DNA, macromolecules, ionic species) and/or colloidal particles with an unprecedented precision \cite{eijkel2005,han2008molecular,napoli2010,bernate2012}.
These devices usually work with single-phase fluid solvents and must combine external electromagnetic fields, electrolyte solutes in proper concentration, and formation of electric double layers in order to induce energy landscapes with directional asymmetry (i.e., ratchet potentials). 
A different class of ratchet systems involving multiphase fluids has been demonstrated to produce ``self-propulsion'' of micro- or millimeter-sized droplets by combining micro/nanostructured surfaces, thermal/chemical gradients, and/or mechanical vibration \cite{mahadevan2004biomimetic,prakash2008surface,noblin2009ratchetlike,mettu2010stochastic,linke2006,quere2013leidenfrost}.
Self-propulsion mechanisms in these multiphase systems are attributed to diverse dynamic phenomena, such as capillarity and contact angle hysteresis \cite{noblin2009ratchetlike,mettu2010stochastic}, or evaporation flows and the Leidenfrost effect \cite{linke2006,quere2013leidenfrost}, where thermal fluctuations play a secondary role.

%
%
%%%%%%%%%%%%%%%%%%%%%%%%%%%%%%%%%%%%%%%%%%%%%%%%%%%%%%%%%%%%%%%%%%%%%%%%%%%%%%%%%%%%%%%%%%%%%%%%%%   
% STUDIED SYSTEM - TERRRACED TOPOLOGY
%%%%%%%%%%%%%%%%%%%%%%%%%%%%%%%%%%%%%%%%%%%%%%%%%%%%%%%%%%%%%%%%%%%%%%%%%%%%%%%%%%%%%%%%%%%%%%%%%%
There is a class of multiphase (two fluid) system that can perform as a thermal ratchet under isothermal and incompressible conditions, with or without the presence of electrolyte solutes and net surface charge. 
In this class of system the thermal ratchet mechanism is enabled by surface nanostructures that induce surface energy barriers with directional asymmetry.
The particular configuration considered in this work, illustrated in Fig.~\ref{fig:1}a, consists of two macroscopically immiscible liquids (fluid-1 and fluid-2) confined in a slit-shaped channel or pore of height $h_o$, length $l$, and width $w\ll h_o$.
The surfaces confining the fluids are chemically homogeneous and neutrally charged. 
One of the surfaces has a terraced structure with regular tread length $s$ and riser height $r$ [cf. Fig.~\ref{fig:1}a] of nanometric dimensions.
Similar terraced structures have been synthesized on crystalline substrates via diverse nanofabrication techniques such as wet etching, high-temperature annealing, and deposition of epitaxial films \cite{rosei2004,kawasaki1994,luttrell2009new,guisinger2009nanometer}.
The studied terraced structure with $N_S$ steps reduces the local height of the channel according to $h(x)=h(0)-r \lfloor x/s \rfloor$ for $0\le x< (N_S+1) s$ (here, $\lfloor x \rfloor\equiv\mathrm{floor}(x)$ is the floor function and $x$ is the coordinate in the longitudinal direction).
In the presence of an interface between two immiscible fluids, the interplay between thermal motion and surface energy barriers induced by the nanoscale structure can drive imbibition and filling/drainage processes in micro/nanoscale channels or pores for a range of wettability conditions unanticipated by conventional wetting models.
%
%
%
%%%%%%%%%%%%%%%%%%%%%%%%%%%%%%%%%%%%%%%%%%%%%%%%%%%%%%%%%%%%%%%%%%%%%%%%%%%%%%%%%%%%%%%%%%%%%%%%%%%%%%%%%%%%%%%%%%
% ANALYTICAL MODEL: SMOLUCHOWSKI AND DAMPING FACTOR
%%%%%%%%%%%%%%%%%%%%%%%%%%%%%%%%%%%%%%%%%%%%%%%%%%%%%%%%%%%%%%%%%%%%%%%%%%%%%%%%%%%%%%%%%%%%%%%%%%%%%%%%%%%%%%%%%%

\section{Theoretical Description: Thermally Driven Wetting}

\begin{figure}
\center
\includegraphics[angle=0,width=0.65\linewidth]{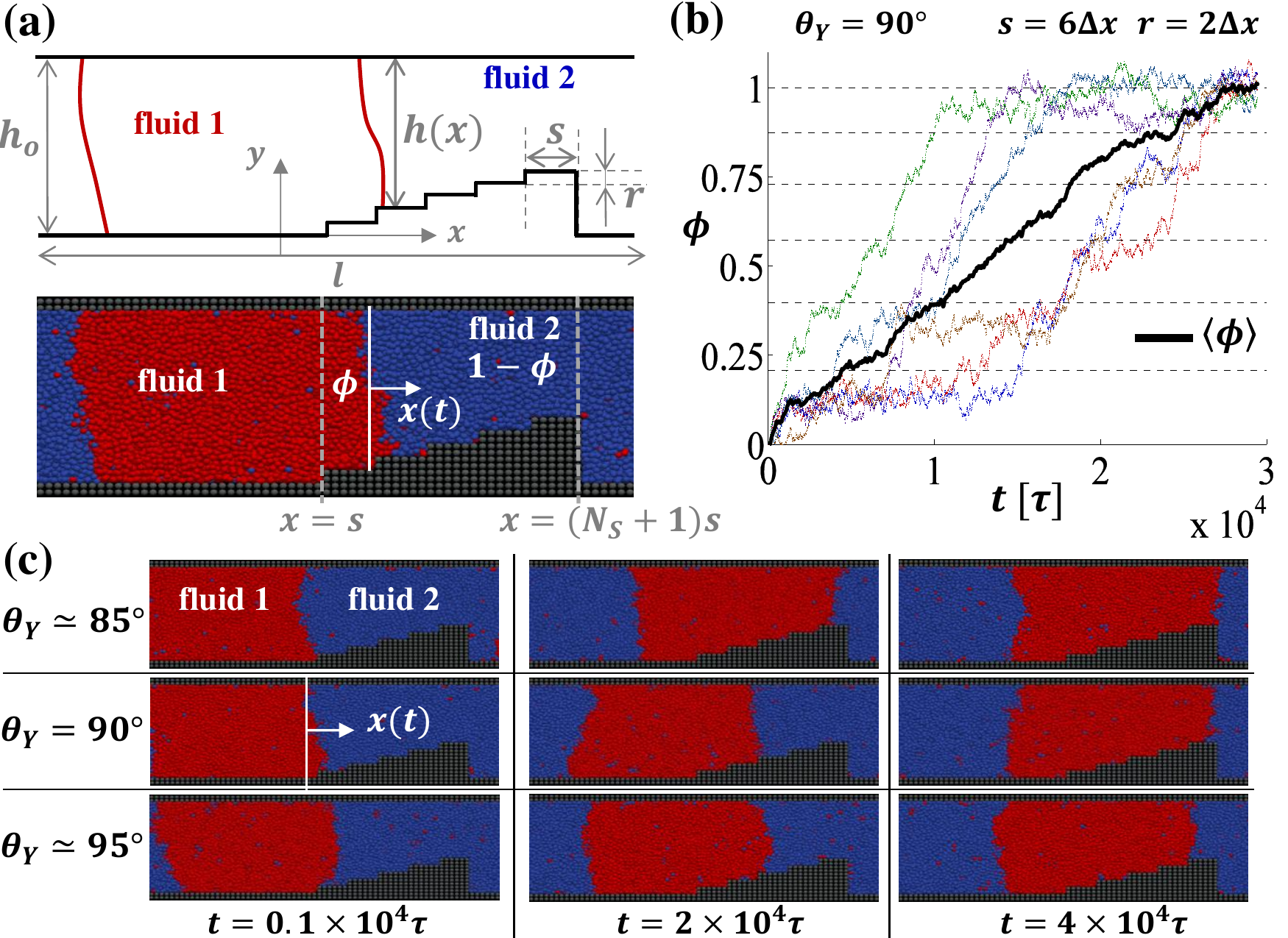}
\vskip -5pt
\caption{Modeled system and geometric configuration.
(a) Slit channel (height $h_o$, length $l$, and width $w$) confining two immiscible liquids. 
A terraced structure (tread $s$ and riser $r$) of nanoscale dimensions lies on the bottom wall.
%
%The height profile $h(x)$ induces surface energy barriers $\Delta U_-=\gamma w r$ at positions $x_n=n s$ ($n=1,N_S$).
%
(b) Volume fraction $\phi(t)$ in different MD realizations and its ensemble average $\langle \phi (t) \rangle$ (time reported in MD units $\tau$~$\simeq$~1~ps).
MD simulations correspond to $s=6\Delta x$ and $r=6\Delta x$ under neutral wetting conditions $\theta_Y=90^\circ$.
Dashed lines indicate filling fractions $\phi_n=\phi(x_n)$ for $x(t)=n s$ ($n=1,N_S+1$).
(c) Sequence of three time instances in MD simulations for contact angles $\theta_Y$~$\simeq$~85--95$^\circ$.
Imbibition of fluid-1 into the volume $V$ above the terraced structure is observed for $\theta_Y\gtrsim 90^\circ$ when capillary forces are neutral or negative $F=-\partial U/\partial x \le 0$.
}
\label{fig:1}
\vskip -10pt
\end{figure}

Analytical descriptions of thermally-driven wetting processes must consider that atoms or molecules in a liquid-fluid interface undergo thermal motion.
We will analyze the case of unidirectional motion described by the average position $x(t)$ of all atoms of the first fluid species (fluid-1) that lie at the front liquid-liquid interface [cf. Fig.~\ref{fig:1}a]. 
Adopting the average interface position to describe the dynamics of the confined molecular fluids implies projecting the (multidimensional) system energy landscape onto a one-dimensional profile $U(x)$ along a ``reaction coordinate'' $x$.
The sequence of random displacements of the front interface position can be statistically described by the conditional probability density $p(x,t)\equiv p(x,t|x_o,t_o)$; here, $x_o=x(t_o)$ is the average interface position observed at a time $t_o$. 
The stationary probability density $p(x,t\to\infty)=Z^{-1} \exp[-U(x)/k_B T]$ is prescribed by the free energy profile $U(x)$ and the thermal energy $k_B T$; here, $Z$ is the corresponding partition function, $k_B$ is the Boltzmann constant and $T=\mathrm{const.}$ is the system temperature.
Assuming overdamped Brownian dynamics, the time evolution of the probability density $p(x,t)$ is governed by the Smoluchowski diffusion equation
\begin{equation}
\frac{\partial}{\partial t}p(x,t)
=\left[\frac{\partial^2}{\partial x^2} \frac{k_B T}{\xi(x)} 
+\frac{\partial}{\partial x} \frac{1}{\xi(x)} \frac{\partial U}{\partial x}\right] p(x,t) 
\label{eq:smoluchowski}
\end{equation}
where $\xi(x)$ is the local friction coefficient or resistivity (i.e., the inverse of the mobility). 
For the studied conditions we consider a linear friction force $F_f=-\xi dx/dt$ that is mainly due to hydrodynamic effects and thus
\begin{equation}
\xi(x)= k_H \mu h(x) w \left[ l- s N_S +\sum_{n=1}^{N_S} \frac{s}{(h_o - n r)^2}\right]
\label{eq:xi}
\end{equation} 
where $k_H$ is a drag coefficient, $\mu$ is the shear viscosity of the confined fluids, and $N_S$ is the total number terraces in the structure.
For analytical simplicity, we consider in Eq.~\ref{eq:xi} the case that both fluid-1 and fluid-2 have the same viscosity $\mu=\mu_1=\mu_2$; expressions for $\mu_1\ne\mu_1$ can be readily derived using similar hydrodynamic arguments.
Analytical estimates of the drag coefficient $k_H$ in Eq.~\ref{eq:xi} can be obtained by making simplifying assumptions about the modes of translation of the interface and the hydrodynamic velocity profiles induced.
A drag coefficient $k_H=4$ is obtained by naively assuming that a linear flow profile  (i.e., Couette flow between two flat surfaces) develops after the sudden displacement of the contact line on one wall, while the contact line on the opposite wall remains stationary. 
%
%
%%%%%%%%%%%%%%%%%%%%%%%%%%%%%%%%%%%%%%%%%%%%%%%%%%%%%%%%%%%%%%%%%%%%%%%%%%%%%%%%%%%%%%%%%%%%%%%%%
% FREE ENERGY
%%%%%%%%%%%%%%%%%%%%%%%%%%%%%%%%%%%%%%%%%%%%%%%%%%%%%%%%%%%%%%%%%%%%%%%%%%%%%%%%%%%%%%%%%%%%%%%%%
\\\indent For isothermal and incompressible conditions and assuming sharp interfaces, the free energy profile is determined by surface energy contributions   
\begin{equation}
U(x)=\gamma [w h(x) - \cos\theta_Y A_{1S}(x)] + \mathrm{const}.
\label{eq:U}
\end{equation}
Here, $\gamma$ is the interfacial energy of the liquid-liquid interface, $\theta_Y$ is the Young contact angle measured on the fluid-1 phase, and $A_{1S}$ is the area of the interface between the fluid-1 and solid phases. 
In the studied configuration the distance between the front and rear liquid-liquid interfaces is larger than the length of the terraced structure, which allows us to simplify the analysis.
While the front interface moves inside each terrace there is a linear change in the interfacial area $A_{1S}(x)$ that results from conservation of volume, and thus we have
\begin{equation}
\frac{\partial U(x)}{\partial x}= -F_n = -2 n \gamma \cos\theta_Y  \frac{w r}{h_o} 
\label{eq:dudx}
\end{equation}
for $n s< x< (n+1) s$ ($n=1,N_S$).
%
%
%%%%%%%%%%%%%%%%%%%%%%%%%%%%%%%%%%%%%%%%%%%%%%%%%%%%%%%%%%%%%%%%%%%%%%%%%%%%%%%%%%%%%%%%%%%%%%%%%
% DIFFUSION SEMI-INFINITE DOMAIN - REFLECTION BOUNDARY
%%%%%%%%%%%%%%%%%%%%%%%%%%%%%%%%%%%%%%%%%%%%%%%%%%%%%%%%%%%%%%%%%%%%%%%%%%%%%%%%%%%%%%%%%%%%%%%%%
\\\indent The energy profile $U(x)$ (Eq.~\ref{eq:U}) exhibits sharp energy increments $\Delta U_{-}\simeq \gamma w r$ when moving in the negative $x$-direction across the edge of a terrace at position $x_n=n s$ ($n=1,N_S$).
Hence, periodic energy barriers $\Delta U_{-}$ at regular steps $s$ hinder backward random displacements as the liquid-liquid interface undergoes thermal motion along the terraced surface.
The time to cross over an energy barrier induced by nanoscale surface features can be predicted via Kramers theory of thermally-activated transitions, as documented in prior work \cite{colosqui2013,razavi2014}.
For the present analysis, it suffices to recognize that a mean time $T_{-}\propto \exp(\Delta U_{-}/k_B T)$ must elapse before observing a backward displacement of the interface over a terrace edge.   
Therefore, over a time interval $t<T_{-}$ the presence of the energy barrier $\Delta U_{-}$ can be treated as a reflective boundary condition for Eq.~\ref{eq:smoluchowski} imposed at the edge of each terrace.
Furthermore, the friction coefficient (Eq.~\ref{eq:xi}) and the wetting or capillary force (Eq.~\ref{eq:dudx}) remain constant within each of the $N_S$ terraces.
Hence, we have $\xi_n=\xi(x)=\mathrm{const}$ and $F_n=-\partial U(x)/\partial x=\mathrm{const}$ for $n s<x< (n+1) s$, which facilitates the analytical solution of Eq.~\ref{eq:smoluchowski}.
In order to solve Eq.~\ref{eq:smoluchowski} within each terrace for which the edge lies at $x_n=n s$, it is convenient to introduce the dimensionless position 
$\overline{x}=(x-x_n)/s$ and time
$\overline{t}=(t-t_o) (k_B T/\xi_n s^2)$.
In addition, we introduce the dimensionless force parameter $C_n=F_n s/k_B T$ given by the ratio of the work performed by the wetting force to the thermal energy within each terrace. 
For a reflective boundary at $x_n$ and initial condition $x_o=x_n$ the analytical solution of Eq.~\ref{eq:smoluchowski} is given by \cite{smoluchowski1916,lamm1983} $p_n(\overline{x},\overline{t})\equiv p(\overline{x},\overline{t};C_n)=\overline{p}_n/s$, where    
\begin{eqnarray}
\label{eq:pn}
\overline{p}_n(\overline{x},\overline{t})= \frac{1}{\sqrt{\pi \overline{t}}} 
\exp\left[-\frac{(\overline{x}+C_n \overline{t})^2}{4\overline{t}} \right]
+ \frac{1}{2} C_n 
\exp\left[-C_n\overline{x}\right]
\mathrm{erfc}\left[\frac{(\overline{x}-C_n \overline{t})}{\sqrt{4\overline{t}}}\right] 
\end{eqnarray}
is a dimensionless function determined by the variables $\overline{x}\ge 0$ and $\overline{t}\ge 0$ for a given force parameter $C_n$.
The mean displacement of the front liquid-liquid interface according to Eq.~\ref{eq:pn} is  $\langle x(t)-x_n \rangle=s {\cal X}_n(\overline{t})$, where 
\begin{equation}
{\cal X}_n(\overline{t})= \int_{0}^\infty \overline{p}_n(\overline{x},\overline{t}) \overline{x} d\overline{x} 
\label{eq:meandx}
\end{equation}
is the dimensionless mean displacement within the $n$-th terrace. 
Taking the upper integration limit to infinity in Eq.~\ref{eq:meandx} is an approximation valid for finite times $t\ll (n-N_S+1)^2 s^2 k_B T/ \xi_n$. 
From Eq.~\ref{eq:meandx} we can estimate the mean time 
\begin{equation}
t_n=t_1+\frac{s^2}{k_B T}\sum_{k=1}^{n-1} \xi_k {\cal X}_k^{-1}(1) 
\label{eq:tn}
\end{equation} 
at which $\langle x(t_n) \rangle=x_n$ ($n=1,N_S+1$). 
Here, ${\cal X}_n^{-1}$ is the inverse of the dimensionless mean displacement function in Eq.~\ref{eq:meandx} and $t_1$ is the time at which $\langle x(t_1)\rangle =s$. 

A few comments aobout the derived expressions are in order.
Analytical integration in Eq.~\ref{eq:meandx} is feasible and thus the dimensionless time 
${\cal T}_n=(t_{n+1}-t_n) k_B T/ \xi_n s^2$  
to traverse the $n$-th terrace ($n=1,N_S$) can be obtained by solving the implicit equation 
${\cal X}_n({\cal T}_n)=1$. 
The inverse function ${\cal X}_n^{-1}(1)={\cal T}_n \equiv {\cal T}(C_n)$ in Eq.~\ref{eq:tn} is uniquely determined by the dimensionless force parameter $C_n=F_n s/k_B T$; the function ${\cal T}_n\to\infty$ diverges at $C_n=-1$ and decays ${\cal T}_n\to 0$ for $C_n\to\infty$. 
An approximate explicit expression ${\cal T}_n=(4/\pi C_n^2)(1-\sqrt{1+(\pi/2)C_n})^2$ can be obtained from a first-order Taylor expansion of Eq.~\ref{eq:pn} about $C_n=0$;  this approximation yields less than 10\% error for $|C_n|<1/2$.
For a neutral contact angle $\theta_Y=90^\circ$ the wetting force vanishes $F_n=0$ and thus ${\cal T}_n=\pi/4$ for $n=1,N_S$. 
Moreover, under neutral wetting conditions Eq.~\ref{eq:meandx} predicts a mean displacement $\langle x(t) \rangle=\sqrt{2 D_o t}$ characteristic of diffusive processes, with an effective diffusivity $D_o=(2/\pi) (k_BT/\xi_n)$.   
Notably, forward liquid displacements $\langle x(t) \rangle>0$ against wetting forces $F_n<0$ are expected to occur for contact angles $\theta_Y > 90^\circ$ provided that 
$-2 n \gamma \cos\theta_Y w \times (r/h_o) < k_B T/s$. 
%
%
%
%%%%%%%%%%%%%%%%%%%%%%%%%%%%%%%%%%%%%%%%%%%%%%%%%%%%%%%%%%%%%%%%%%%%%%%%%%%%%%%%%%%%%%%%%%%%%%%%%%   
% MOLECULAR DYNAMICS
%%%%%%%%%%%%%%%%%%%%%%%%%%%%%%%%%%%%%%%%%%%%%%%%%%%%%%%%%%%%%%%%%%%%%%%%%%%%%%%%%%%%%%%%%%%%%%%%%%
%
\section{Molecular Dynamics Simulations}
%
%%%%%%%%%%%%%%%%%%%%%%%%%%%%%%%%%%%%%%%%%%%%%%%%%%%%%%%%%%%%%%%%%%%%%%%%%%%%%%%%%%%%%%%%%%%%%%%%%%%%%%%%%%%%%%%%%%%%%%%%%%%%%%%%
% Figure1: Problem setup
%%%%%%%%%%%%%%%%%%%%%%%%%%%%%%%%%%%%%%%%%%%%%%%%%%%%%%%%%%%%%%%%%%%%%%%%%%%%%%%%%%%%%%%%%%%%%%%%%%%%%%%%%%%%%%%%%%%%%%%%%%%%%%%%
%%%%%%%%%%%%%%%%%%%%%%%%%%%%%%%%%%%%%%%%%%%%%%%%%%%%%%%%%%%%%%%%%%%%%%%%%%%%%%%%%%%%%%%%%%%%%%%%%%%%%%%%%%%%%%%%%%%%%%%%%%%%%%%%
%
%
%
In order to verify analytical predictions and underlying assumptions we perform fully atomistic molecular dynamics (MD) to simulate the dynamics arising from couplings between Brownian motion, hydrodynamic effects, and wetting forces. 
The MD techniques employed in this work are extensively described in the literature \cite{rapaport,frenkel2002,koplik1989,koplik1995} and prior work by the authors \cite{colosqui2013,razavi2014}.
The simulated system [see Fig.~1a] comprises two monatomic liquids (atomic species $i=1,2$) labeled as fluid-1 and fluid-2, and a crystalline solid (atomic species $i=3$).
Atomic interactions are modeled by pairwise Lennard-Jones potentials
 $u_{LJ}(r)=\varepsilon[(r/\sigma)^{-12}-c_{ij}(r/\sigma)^{-6}]$, 
where $\varepsilon$ is a characteristic interaction energy, $c_{ij}=c_{ji}$ is a symmetric attraction coefficient between species ($i,j=1,3$), $r$ is the interatomic distance between any two atoms, and $\sigma$ is the diameter of the repulsive core, which roughly correspond to the atomic diameter.
Following conventional techniques for computational efficiency, atomic interactions are not computed for $r> 2.5 \sigma$. 
The time scale of atomic displacements is $\tau=\sigma\sqrt{m/\varepsilon}$ and is in the order of picoseconds for simple molecular liquids; the atomic mass $m$ of the liquid species is set to be equal for both modeled liquids.  
The equations of motion are integrated using a fifth-order prediction-correction algorithm with a time step $\delta=0.004\tau$. 
A Nose-Hoover thermostat \cite{rapaport,frenkel2002,nose1984} models the interaction with a thermal bath, regulating the system temperature to a prescribed value.
In all MD simulations in this work the prescribed temperature is $T=2 \varepsilon/k_B T$, the mean number density is $\langle\rho\rangle=0.8/\sigma^3$ and the shear viscosity is $\mu=2.4 m/(\sigma \tau)$ for both liquids. 
Solid atoms form a face-cubic-centered (fcc) lattice with uniform spacing $\Delta x= 0.8^{-1/3}\sigma$; typical values of $\Delta x$ for crystalline solids range between 0.2 and 0.5 nm. 
The set of values employed for the attraction coefficients ($c_{12}=0.5$, $c_{13}=0.8$, $0.75 \le c_{23}\le 0.95$, and  $c_{ii}=1$) render two macroscopically immiscible fluids with a liquid-liquid interfacial tension $\gamma=1.2 \varepsilon/\sigma^2$ and Young contact angles near neutral wetting conditions $85^\circ\le\theta_Y\le 100^\circ$.
%
%

%%%%%%%%%%%%%%%%%%%%%%%%%%%%%%%%%%%%%%%%%%%%%%%%%%%%%%%%%%%%%%%%%%%%%
% RESULTS DESCRIPTION
%%%%%%%%%%%%%%%%%%%%%%%%%%%%%%%%%%%%%%%%%%%%%%%%%%%%%%%%%%%%%%%%%%%%%
\section{Results: Theory and Molecular Dynamics }
%
%%%%%%%%%%%%%%%%%%%%%%%%%%%%%%%%%%%%%%%%%%%%%%%%%%%%%%%%%%%%%%%%%%%%%%%%%%%%%%%%%%%%%%%%%%%%%%%%%%%%%%%%%%%%%%%%%%
% TERRACED STRUCTURE GEOMETRY & FILLING DYNAMICS
%%%%%%%%%%%%%%%%%%%%%%%%%%%%%%%%%%%%%%%%%%%%%%%%%%%%%%%%%%%%%%%%%%%%%%%%%%%%%%%%%%%%%%%%%%%%%%%%%%%%%%%%%%%%%%%%%%
%
The studied geometric configuration [cf. Fig.~\ref{fig:1}a] consists of a slit nanoscale channel of height $h_o$~=~30--60$\Delta x$, length $l$~=~70$\Delta x$, and width $w$~=~10$\Delta x$; periodic boundary conditions are applied in the $x$- and $z$-direction.
The bottom wall has different surface structures with $N_S$~=~5--11 terraces of length $s$~=~3--9$\Delta x$ and riser height $r$~=~1--2$\Delta x$. 
The volume confined above the terraced structure is 
$V=w s \sum_{k=1}^{N_S} [h_o - k r]$. 
The fluid-1 phase fills a fraction 
$\phi(x)= x w(h_o+s)/V  - x^2(wr/2Vs)+{\cal O}(r/h(x))$ of the volume $V$ as the front liquid interface moves within $s\le x(t) \le (N_S+1) s$ [cf. Fig.~\ref{fig:1}a]. 
In our MD simulations, the number density is constant and the filling fraction $\phi(t)=n_1/(n_1+n_2)$ is directly computed from the number of atoms of fluid-1, $n_1$, and fluid-2, $n_2$, occupying the volume $V$.
At initialization, the volume $V$ is fully occupied by the fluid-2 phase and thus $\phi(t=0)=0$.

The filling fraction $\phi(t)$ for six MD realizations 
\footnote[1]{Reported average quantities correspond to ensemble average over 10 to 12 MD simulations for the same initial macroscopic conditions and different initial conditions for the atomic velocities.}
and the ensemble average $\langle \phi(t) \rangle$ are reported in Fig.~\ref{fig:1}b for one of the studied structures ($s=6\Delta x$, $r=2\Delta x$) and neutral wetting conditions ($c_{13}=c_{23}=0.8$).
As seen in Figs.~\ref{fig:1}b, the mean filling fraction $\langle \phi(t) \rangle$ uniformly increases with time while individual MD realizations exhibit rapid transitions between ``long-lived'' metastable states at specific values $\phi_n\simeq\phi(x_n)$ ($n=1,N_S+1$).   
The imbibition process occurs within a range of Young contact angles $\theta_Y \gtrsim 90$ [cf. Fig~\ref{fig:1}c] for which capillary forces can be neutral or negative ($F_n=-\partial U/\partial x\lesssim 0$).
Displacement beyond the last terrace edge, where $\phi(x_{N_S+1})=1$, is prevented by a large energy barrier 
$\Delta U_+=\gamma N_S r w$.
% 
%

%%%%%%%%%%%%%%%%%%%%%%%%%%%%%%%%%%%%%%%%%%%%%%%%%%%%%%%%%%%%%%%%%%%%%%%%%%%%%%%%%%%%%%%%%%%%%%%%%%%%%%%%%%%%%%%%%%%%%%%%%%%%%%%%
% Figure2: MD vs. LD vs theory
%%%%%%%%%%%%%%%%%%%%%%%%%%%%%%%%%%%%%%%%%%%%%%%%%%%%%%%%%%%%%%%%%%%%%%%%%%%%%%%%%%%%%%%%%%%%%%%%%%%%%%%%%%%%%%%%%%%%%%%%%%%%%%%%
\begin{figure}
\center
\includegraphics[angle=0,width=0.65\linewidth]{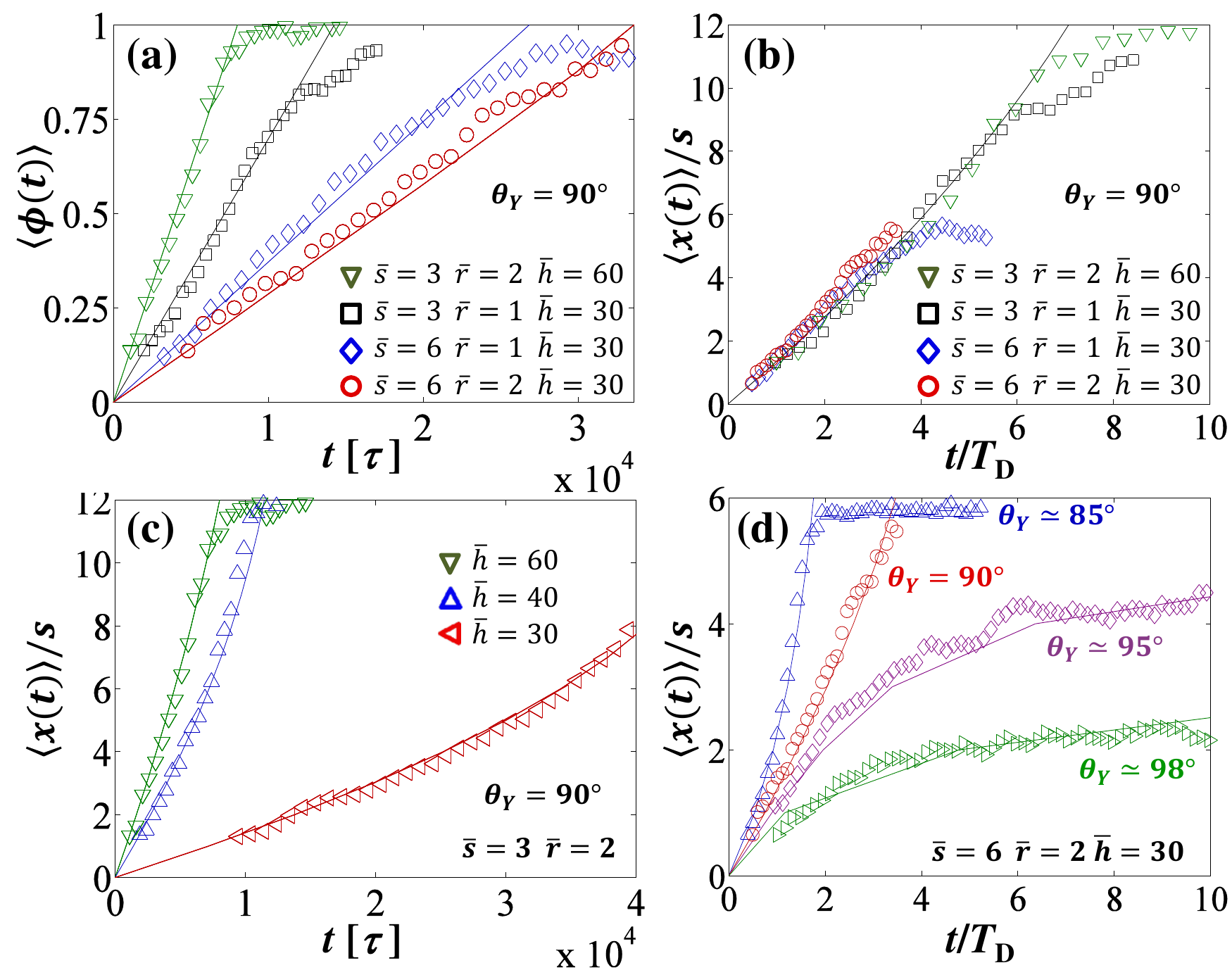}
\vskip -5pt
\caption{Thermally-driven wetting on terraced nanostructures.
Solid lines: linear interpolation between coordinate pairs ($x_n$,$t_n$); $x_n=n s$ ($n=1,N_S+1$) and $t_n$ given by Eq.~\ref{eq:tn}.
Markers: ensemble average quantities from MD simulations [see legends].
Geometric dimensions are in lattice units: channel height $\overline{h}=h_o/\Delta x$~=~30--60, terrace length $\overline{s}=s/\Delta x$~=~3--6, riser height $\overline{r}=s/\Delta x$~=~1--2.
(a)--(c) Imbibition under neutral wetting $\theta_Y=90^\circ$:
(a) mean filling fraction $\langle \phi(t) \rangle$ (time in MD units $\tau$);
(b) dimensionless displacement $\langle x(t/T_D)\rangle/s$ where $T_D=T_D=s^2 \xi_o/k_B T$;
(c) dimensionless displacement $\langle x(t)\rangle/s$ for different channel heights.
(d) Dimensionless displacement $\langle x(t/T_D)\rangle/s$ for contact angles $\theta_Y$~=~85$^\circ$, 90$^\circ$, 95$^\circ$, and 98$^\circ$ for which $C_{N_S}$~=~0.7, 0, -0.7, and -1, respectively. 
}
\vskip -20pt
\label{fig:2}
\end{figure}
%%%%%%%%%%%%%%%%%%%%%%%%%%%%%%%%%%%%%%%%%%%%%%%%%%%%%%%%%%%%%%%%%%%%%%%%%%%%%%%%%%%%%%%%%%%%%%%%%%%%%%%%%%%%%%%%%%%%%%%%%%%%%%%%
%
%
%

\begin{figure}
\center
\includegraphics[angle=0,width=0.65\linewidth]{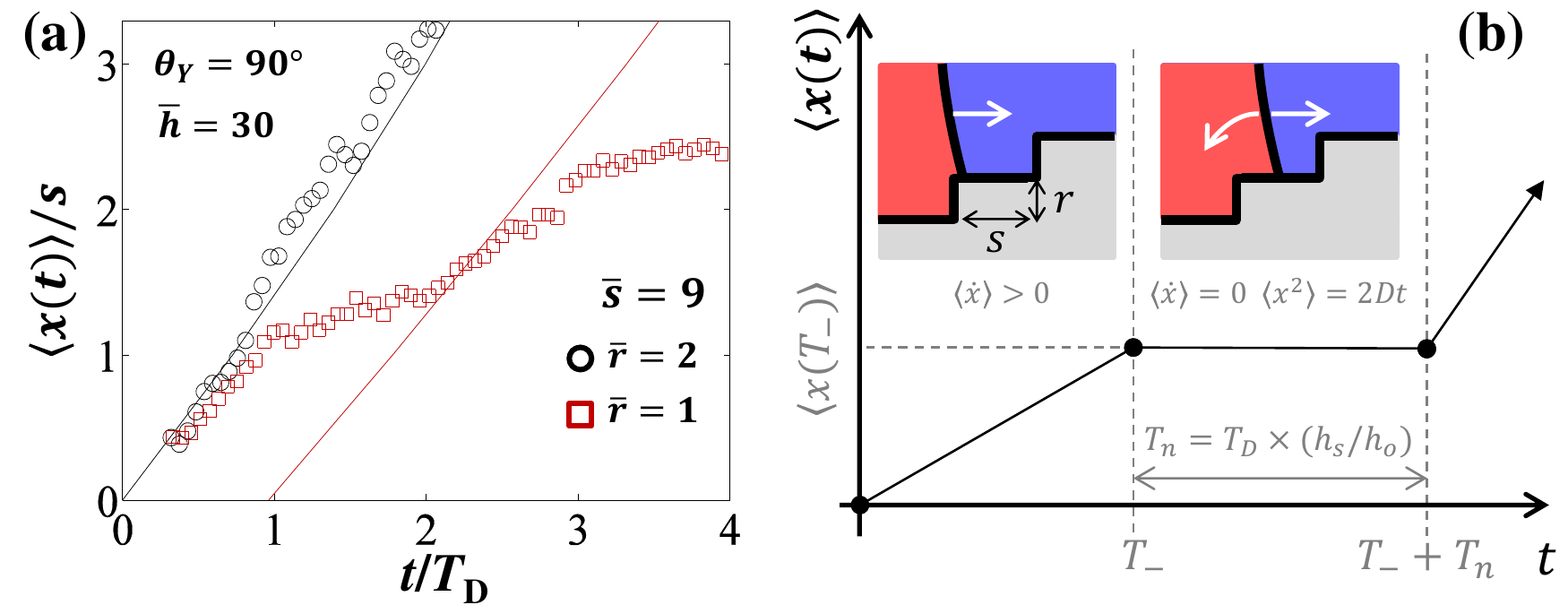}
\vskip -10pt
\caption{Failure of the thermal ratchet mechanism. 
(a) Dimensionless displacement $\langle x(t/T_D)\rangle/s$ in MD simulations for large terrace length $\overline{s}=9$ and two riser heights $\overline{s}$~=1--2  ($\theta_Y$~=~90$^\circ$, $\overline{h}=30$).
(b) Schematic of the failure and onset of diffusive motion with zero mean displacement for $T_-\le t <T_-+T_n$.
}
\vskip -20pt
\label{fig:3}
\end{figure}

Theoretical predictions from Eq.~\ref{eq:tn} for mean displacements, $\langle x(t_n) \rangle=x_n$, and filling fractions, $\langle \phi(t_n) \rangle=\phi(x_n)$, are in close agreement with MD simulations reported in Figs.~\ref{fig:2}--\ref{fig:3} when using $k_H$~=~4--5.5 in Eq.~\ref{eq:xi}  for different structures.
As seen in Fig.~\ref{fig:2}a, the mean filling rate is approximately constant 
$\langle \dot{\phi}(t) \rangle \simeq \mathrm{const}$ under neutral wetting conditions $\theta=90^\circ$ for the different studied structures.
Notably, mean displacements $\langle x(t) \rangle$ collapse to a unique curve when scaling by the terrace length $s$ and diffusion time $T_D=s^2 \xi_o/k_B T$ as showed in Fig.~\ref{fig:2}b; here, $\xi_o=\xi(0)$ is the resistivity in Eq.~\ref{eq:xi} for $h(x=0)=h_o$. 
As shown in Fig.~\ref{fig:2}c, increasing the channel height $h_o$ for a given length $l$ enhances the interface displacement rate by decreasing the local resistivity $\xi(x)$.
The effect of varying the contact angle $\theta_Y$ and thus the wetting force $F_n$ (Eq.~\ref{eq:tn}) can be seen in Fig.~\ref{fig:2}d.
For $\theta_Y \lesssim 90^\circ$ the mean displacement is enhanced by positive wetting forces $F_n>0$.
As predicted, positive displacements still can be observed for $\theta_Y \gtrsim 90^\circ$ when $F_n s/k_B T >-1$; the most adverse wetting force $F_{N_S} \simeq - k_B T/s$ corresponds to $\theta_Y\simeq 98^\circ$ 
($\overline{s}=s/\Delta x=6$, $\overline{r}=r/\Delta x=2$) in Fig.~\ref{fig:2}d.     
%
%
%
%%%%%%%%%%%%%%%%%%%%%%%%%%%%%%%%%%%%%%%%%%%%%%%%%%%%%%%%%%%%%%%%%%%%%%%%%%%%%%%%%%%%%%%%%%%%%%%%%%%%%%%%%%%%%%%%%%%%%%%%%%%%%%%%
% Figure3: CROSSOVERS
%%%%%%%%%%%%%%%%%%%%%%%%%%%%%%%%%%%%%%%%%%%%%%%%%%%%%%%%%%%%%%%%%%%%%%%%%%%%%%%%%%%%%%%%%%%%%%%%%%%%%%%%%%%%%%%%%%%%%%%%%%%%%%%%
%%%%%%%%%%%%%%%%%%%%%%%%%%%%%%%%%%%%%%%%%%%%%%%%%%%%%%%%%%%%%%%%%%%%%%%%%%%%%%%%%%%%%%%%%%%%%%%%%%%%%%%%%%%%%%%%%%%%%%%%%%%%%%%%
%
%
%

%%%%%%%%%%%%%%%%%%%%%%%%%%%%%%%%%%%%%%%%%%%%%%%%%%%%%%%%%%%%%%%%%%
%  FAILURE MECHANISM & CONNECTION W/EXPERIMENTAL SYSTEMS
%%%%%%%%%%%%%%%%%%%%%%%%%%%%%%%%%%%%%%%%%%%%%%%%%%%%%%%%%%%%%%%%%%
Results reported in Fig.~\ref{fig:2} confirm that a thermal ratchet mechanism mediated by surface energy barriers $\Delta U_{-}\simeq \gamma w r$ can lead to thermally-driven transport of immiscible liquids in slit channels or pores.
This phenomenon is expected to occur provided that the time $T_{-}\propto \exp(-\gamma w r/k_B T)$ to cross over each energy barrier
is larger than the time $t_n-t_{n-1}$ (Eq.~\ref{eq:tn}) to traverse the $n$-th terrace.
Indeed, MD simulations for large terrace length $s=9\Delta x$ and small riser height $r=\Delta x$ [see Fig~\ref{fig:3}a] report a decay in the mean displacement rate after a time $T_-\simeq T_D$.
Increasing the energy barrier $\Delta U_-$ by increasing the riser height to $r=2\Delta x$ prevents the observed decay in the displacement rate [see Fig~\ref{fig:3}a].
As illustrated in Fig~\ref{fig:3}b, the mean displacement rate is expected to vanish $\langle \dot{x}(t) \rangle\to 0$ for $t>T_-$, after which unbiased Brownian motion persists over a time 
$T_n=s^2 \xi_n/k_B T=T_D\times(h_n/h_o)$.
For $T_-< t \le T_-+T_n$ Brownian motion with a mean square displacement $\langle x^2(t)\rangle = 2 (k_B T/\xi_n) t$ causes the liquid-liquid interface to diffuse to the next terrace edge, after which thermal motion becomes biased, $\langle \dot{x}(t) \rangle >0$, and the cycle repeats. 
%
%%%%%%%%%%%%%%%%%%%%%%%%%%%%%%%%%%%%%%%%%%%%%%%%%%%%%%%%%%%%%%%%%%%%%%%%%%%%%%%%
% Conclusions
%%%%%%%%%%%%%%%%%%%%%%%%%%%%%%%%%%%%%%%%%%%%%%%%%%%%%%%%%%%%%%%%%%%%%%%%%%%%%%%%
\section{Conclusions}
Theoretical description in the framework of statistical physics predicts that surface nanostructures with directional asymmetry can induce nontrivial wetting processes that are beyond the reach of conventional continuum-based models (e.g., Lucas-Washburn equation). 
Fully-atomistic simulations showing close agreement with theoretical predictions document the thermally-induced displacement of immiscible fluids confined in a slit channel or pore with a nanoscale terraced structure.
For the studied nanoscale systems under neutral wetting conditions, a water-air interface would exhibit mean displacement rates between 0.1 and 1 m/s.
Moreover, the observed fluid displacement can oppose the action of wetting or capillary forces.
Hence, the studied mechanism can induce the wetting and dewetting of nanostructured pores or capillaries under unexpected wettability conditions.   
The maximum contact angle for which thermally driven imbibition or drainage occurs against the action of capillary forces is largely determined by the geomtry of the channel and terraced structure.

The analytical approach presented in this work indicates the necessity to consider the thermal motion of liquid-fluid interfaces in order to predict novel phenomena and associated useful effects.
The proposed thermal ratchet mechanism enabled by engineered surface nanostructures could enable novel nanofluidic devices for passive handling and separation.  
Micro/nanostructured surfaces for superhydrophobic or superoleophobic behavior could exploit the thermally driven drainage of micropores and cavities to enhance their performance.
In the presence of electrolyte solutes, the studied thermal ratchet mechanism could be employed to direct the transport of charges, which can potentially enable microfluidic applications for energy harvesting.

%%%%%%%%%%%%%%%%%%%%%%%%%%%%%%%%%%%%%%%%%%%%%%%%%%%%%%%%%%%%%%
%%%%%%%%%%%%%%%%%%%%%%%%%%%%%%%%%%%%%%%%%%%%%%%%%%%%%%%%%%%%%%
% ACKNOWLEDGEMENTS
%%%%%%%%%%%%%%%%%%%%%%%%%%%%%%%%%%%%%%%%%%%%%%%%%%%%%%%%%%%%%%
%%%%%%%%%%%%%%%%%%%%%%%%%%%%%%%%%%%%%%%%%%%%%%%%%%%%%%%%%%%%%%
%
\begin{acknowledgments}
We thank A. Checco and M. Sbragaglia for useful discussions. 
We acknowledge support from the SEED Grant Program by Brookhaven National Laboratory (BNL) and Stony Brook University. 
This work employed computational resources at the BNL Center for Functional Nanomaterials supported by The U.S. DOE under Contract No. DE-SC0012704.
\end{acknowledgments}
%
%
%
%\newpage 
%

\end{document}